\documentclass[journal]{IEEEtran}

\usepackage{amssymb}
\usepackage[cmex10]{amsmath}
\usepackage{stfloats}
\usepackage{graphicx}
\usepackage{subfigure}
\usepackage{tabularx}
\usepackage{epsfig,epsf,color,balance,cite}
\usepackage{verbatim}
\usepackage{url}
\usepackage{bm}

\newtheorem{theorem}{\bf Theorem}

\usepackage{algorithm}
\usepackage{algorithmic}

\usepackage{color}
\definecolor{myc1}{rgb}{0,0,0}
\definecolor{myc2}{rgb}{0,0,0}

\definecolor{myc3}{rgb}{0,0,1}
\begin{document}

%
\title{Energy Efficient Reconfigurable Intelligent Surface Enabled Mobile Edge Computing Networks with NOMA}
%
%
%

\author{
	Zhiyang~Li,~\IEEEmembership{Student~Member,~IEEE,}
	Ming~Chen,~\IEEEmembership{Member,~IEEE,}
        Zhaohui~Yang,~\IEEEmembership{Member,~IEEE,}
        Jingwen~Zhao,~\IEEEmembership{Student~Member,~IEEE,}
        Yinlu~Wang,~\IEEEmembership{Student~Member,~IEEE,}
        Jianfeng~Shi,~\IEEEmembership{Member,~IEEE,}
        and~Chongwen~Huang,~\IEEEmembership{Member,~IEEE.}

\thanks{This work was supported in part by the National Natural Science Foundation of China under Grant 61871128, Grant 61960206006, and Grant 61960206005, in part by the open research fund of National Mobile Communications Research Laboratory, Southeast University (No.2021D11), Universities Natural Science Research Project of Jiangsu Province (20KJB510037). The work of Prof. Huang was supported by the Fundamental Research Funds for the Central Universities. (Corresponding authors: Zhiyang Li, Zhaohui Yang.)}

\thanks{Zhiyang~Li, Ming~Chen, Jingwen~Zhao and Yinlu~Wang are with National Mobile Communications Research Laboratory, Southeast University, Nanjing 211111, China. emails: \{lizhiyang, chenming, zhaojingwen,  yinluwang\}@seu.edu.cn.}

\thanks{Zhaohui Yang is with Centre for
	Telecommunications Research, Department of Informatics Engineering, King’s College London, London WC2B
	4BG, UK. email: yang.zhaohui@kcl.ac.uk.}

\thanks{J. Shi is with the School of Electronic and Information Engineering, Nanjing University of Information Science and Technology, Nanjing, 210044, China; and also with the National Mobile Communications Research Laboratory, Southeast University, Nanjing, 210096, China. email: jianfeng.shi@nuist.edu.cn.}

\thanks{C. Huang is with College of Information Science and Electronic Engineering, Zhejiang University, Hangzhou 310027, China, and with International Joint Innovation Center, Zhejiang University, Haining 314400, China, and also with Zhejiang Provincial Key Laboratory of Info. Proc., Commun. \& Netw. (IPCAN), Hangzhou 310027, China. email: chongwenhuang@zju.edu.cn.}

}
%
%

{}
%



\maketitle

\begin{abstract}
Reconfigurable intelligent surface (RIS) has emerged as a promising technology for achieving high spectrum and energy efficiency in future wireless communication networks. In this paper, we investigate an RIS-aided single-cell multi-user mobile edge computing (MEC) system where an RIS is deployed to support the communication between a base station (BS) equipped with MEC servers and multiple single-antenna users. To utilize the scarce frequency resource efficiently, we assume that users communicate with BS based on a non-orthogonal multiple access (NOMA) protocol. Each user has a computation task which can be computed locally or partially/fully offloaded to the BS. We aim to minimize the sum energy consumption of all users by jointly optimizing the passive phase shifters, the size of transmission data, transmission rate, power control, transmission time and the decoding order. Since the resulting problem is non-convex, we use the block coordinate descent method to alternately optimize two separated subproblems. More specifically, we use the dual method to tackle a subproblem with given phase shift and obtain the closed-form solution; and then we utilize penalty method to solve another subproblem for given power control. Moreover, in order to demonstrate the effectiveness of our proposed algorithm, we propose three benchmark schemes: the time-division multiple access (TDMA)-MEC scheme, the full local computing scheme and the full offloading scheme. We use an alternating 1-D search method and the dual method that can solve the TDMA-based transmission problem well. Numerical results demonstrate that the proposed scheme can increase the energy efficiency and achieve significant performance gains over the three benchmark schemes.
\end{abstract}


%
\IEEEpeerreviewmaketitle

\section{Introduction}
\subsection{Related Works}
%
%
%
%
\IEEEPARstart In the forthcoming sixth-generation (6G) networks, mobile data will experience a phenomenal growth, most of which are generated by edge devices in real-time, such as mobile phones, computers and sensors. Since most of the users are sensors with limited computing capabilities, mobile edge computing (MEC) as a promising technology can efficiently guarantee the computing resource, reduce the capital cost and provide flexibility. In the MEC framework, wireless devices offload their computation intensive or delay-sensitive  tasks to nearby base stations (BSs) or servers at the edge of radio access networks to save the battery power and computation resources\cite{Survey_A,LiH,chen2021communication}.
Driven by the benefits of MEC networks, there were a large number of contributions studying MEC-based systems and numerous efficient algorithms were proposed to solve joint communication and computation resource allocation problems \cite{cdma,Cao}.


Meanwhile, reconfigurable intelligent surface (RIS) can be used as a passive reflector which does not need radio frequency (RF) module to forward signals \cite{CHuang1,yang2020energyefficient,yang2021federated}. The RIS could change the channel condition for the MEC service and significantly increase spectrum spatial efficiency.
The RIS consists of many low-cost and re-configurable printed dipoles, and the smart controller can change the specific angles of these passive elements according to the requirements of systems \cite{CHuang2,wuIRS}. By optimizing the phase shifts of RIS, one can both suppress the interference and boost the desired signal, which enhances transmission security and efficiency\cite{hongartificial, LPan,yang2020RIS}.
\cite{qqq} introduced the structure of RIS that it is a massive low-cost reflecting elements mounted surface, and how the RIS achieved 3-D passive beamforming via the amplitude and phase shift optimization.	
The authors in \cite{SZhangCapacity} introduced the fundamental capacity limit characterization of RIS-aided multiple-input multiple-output (MIMO) communication networks by jointly optimizing the transmit covariance matrix and RIS reflection coefficients.
In \cite{JSCPan}, the RIS was shown to extend the operational range of the wireless powered sensors in simultaneous wireless information and power transfer (SWIPT) networks.
The RIS was shown to significantly enhance the data rate of the cell-edge users in \cite{TCOMpan} for the multicell networks.
By utilizing the benefits of RIS, including improving channel conditions, enhancing reflection-aided beamforming gain and energy transmission, the potential of MEC systems can be significantly improved.
For example, \cite{Latencybai} investigated RIS-based MEC systems, where users could choose to offload partial of their tasks to the MEC with the help of RIS. Numerical results showed that the overall system latency decreases significantly compared with the system without RIS. In \cite{bai2020resource}, the authors considered a wireless powered MEC network and an RIS was used to aid wireless energy transfer and computation offloading. {However, in \cite{Latencybai} and \cite{bai2020resource}, the authors only studied the performance improvement when integrating RIS into MEC networks, without studying the performance under various multiple access schemes.}

On the other hand, non-orthogonal multiple access (NOMA) is recognized as an essential radio access technology in the fifth generation (5G) and beyond 5G \cite{NOMA}.
For traditional orthogonal multiple access (OMA) techniques, such as time division multiple access (TDMA) and orthogonal frequency division multiple access (OFDMA), each resource block is occupied by only one user. With NOMA, users can reuse the same frequency resource to save the energy and spectrum resources.
Since NOMA technique can significantly enhance the system spectrum efficiency, it has been extensively studied in MEC-based systems.
In \cite{NOMAmecDing}, the authors considered a NOMA-assisted MEC system and obtained closed-form expressions, where the energy consumption and the time interval were jointly optimized. \cite{NOMAmecWang} studied a multiuser MEC system where users offload their computation tasks through NOMA technique and the BS decoded all offloaded tasks by successive interference cancellation (SIC), in which the NOMA decoding order was optimized.
\cite{YeY} proposed a hybrid offloading scheme where the users were divided into two groups and their locations followed a homogeneous Poisson Point Process. They also investigated distance-dependent NOMA grouping schemes in MEC networks and used numerical results to show the advantages and disadvantages of each scheme.
The advantages of NOMA technology and its wide applications inspire the idea of using NOMA as a means of transmission in our model.

{When the channel gain differences of users are large or the frequency resource is limited, the advantages of NOMA are more obvious, and RIS can be deployed to help achieve this goal by carefully designing its phase shifts. A design of the RIS-assisted NOMA downlink transmission with the impact of hardware impairments was proposed in \cite{dingz}. Moreover, in \cite{zhengNOMARIS}, the authors considered an RIS assisted NOMA network where a robust beamforming scheme was proposed in an eavesdropping scenario with imperfect channel state information (CSI) for the eavesdropper. Furthermore, multi-cluster multiple-input-single-output (MISO) NOMA communication networks were discussed in \cite{yiqingNOMARIS}, and RISs were utilized to assist the communication. However, the above works \cite{dingz,zhengNOMARIS,yiqingNOMARIS} did not consider RIS-aided NOMA communication for the MEC networks.}

\subsection{Motivations and Contributions}
{The above-mentioned works [14]-[22] mainly studied the NOMA-MEC networks or the RIS assisted MEC networks. Since RIS can significantly enhance the system performance by reconfiguring the channel conditions and NOMA can be used to serve multiple users simultaneously at the same resource block \cite{dingz, zhengNOMARIS, yiqingNOMARIS}, we consider to utilize RIS in the NOMA-MEC networks to further improve the energy efficiency. In this paper, we focus on an RIS-aided NOMA-MEC network where the direct links between users and the BS are in poor conditions and the RIS is used to guarantee the quality of service (QoS) requirements of  MEC services. Our contributions can be summarized as follows:
\begin{itemize}
	\item  We propose an RIS-aided multi-user MEC MISO system, where a multi-antenna BS (also acts as an MEC server) serves multiple single-antenna users using NOMA. To minimize the sum energy consumption of all users, we jointly optimize the passive reflecting beamforming (i.e., phase shifts of RIS), the number of transmission data, transmission rate, power control, transmission time as well as the decoding order.
	\item To tackle the formulated non-convex problem, we propose a low-complexity algorithm by solving two subproblems iteratively. In particular, for the first subproblem with given phase shifts, we apply the dual decomposition method to obtain the optimal closed-form solution as well as the optimal decoding order. Then for the second subproblem with given power control, we utilize the penalty method to transform this subproblem into a series of convex problems and obtain a suboptimal solution.
	\item  To demonstrate the effectiveness of our proposed scheme, we provide a TDMA-based transmission model as a benchmark. To solve the energy efficient resource allocation for TDMA case, we propose the dual decomposition method with 1-D search.
	\item Simulation results show that our proposed RIS-MEC with NOMA scheme can dramatically decrease the sum energy consumption at users compared with three conventional benchmark schemes. In addition, we discuss the impact of distances on the performance of our proposed system.
\end{itemize}}

\subsection{Organization and Notation}
The rest of this paper is organized as follows. In Section II, we introduce the system model and problem formulation of the RIS-aided NOMA-based MEC network. In Section III, we introduce an efficient block coordinate descent algorithm for solving the proposed problem. In Section IV, we propose a TDMA-based partial offloading scheme as a benchmark and solve the TDMA-based transmission problem. In Section V, we show the numerical results and evaluate the performance of our proposed schemes. Finally, we give the conclusions in Section VI.

\textit{Notations:} Scalars, vectors and matrices are denoted by italic letters, bold lower and upper case letters, respectively. $\mathbb R^{x \times y}$ and $\mathbb C^{x \times y}$ denote the space of $x \times y$ real-valued and complex-valued matrices, respectively. $\mathbb C^x$ denotes the sets of all complex-valued $x \times 1$ vector. $\text {Re}(x)$ denotes the real part of a complex number $x$. $\|\pmb x\|$ denotes the Euclidean norm of the complex-valued vector $\pmb x$ and diag$(\pmb x)$ denotes the diagonal matrix and the diagonal elements corresponding to the elements in vector $\pmb x$. The distribution of a circularly symmetric complex Gaussian (CSCG) random vector with mean vector $\mu$ and covariance matrix $\sigma^2$ is denoted by $\pmb n \sim \mathcal C \mathcal N(\pmb \mu, \pmb \sigma^2)$, where `$\sim$' means `distributed as'.
\begin{figure}[htb]
	\centering
	\includegraphics[width=0.5\textwidth]{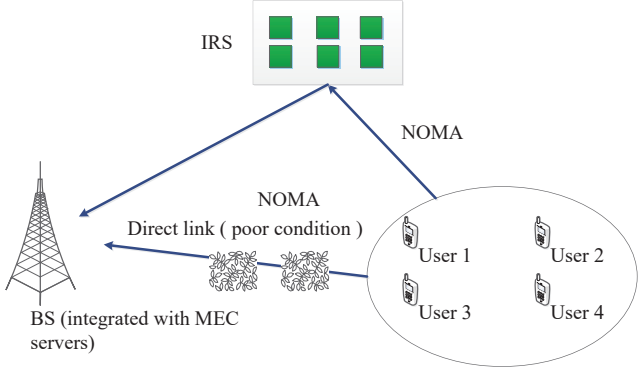}
	\vspace{-1.5em}
	\caption{RIS-based MEC uplink system model.}\label{fig1}
	\vspace{-.5em}
\end{figure}

\section{System Model and Problem Formulation}
\subsection{System Model}
As shown in Fig.1, we consider an RIS-enhanced uplink system with one BS, $K$ users and one RIS, which helps the communication between the BS and users.
The number of transmit antennas at the BS and that of reflecting elements at the RIS are respectively denoted by $M$ and $N$.
Let $\mathcal K=\{1, 2, \cdots, K\}$ denote the set of all users.
The BS schedules the users to completely or partially offload tasks.
The users with partial or complete offloading respectively offload a fraction of or all input data to the BS, while the users with partial or no offloading respectively compute a fraction of or all input data using local central processing unit (CPU).
Due to the small latency of cloud computing and small sizes of computation results, the time of cloud computing and downloading from the BS is negligible compared to the time of mobile offloading and local computing.

\subsection{Task Computing Model}
The local computing model is described as follows.
Since only $d_{k}$ bits are offloaded to the BS, the remaining $R_{k}-d_{k}$ bits are needed to be computed locally at user $k$.
The total CPU cycles for local computation at user $k$ is given by
\begin{equation}\label{sys1eq1}
\mathcal C_{k}^{\text{Loc}} = (R_{k}-d_{k}) C_{k}, \quad \forall k\in\mathcal K,
\end{equation}
where $C_{k}$ is the number of CPU cycles required for computing 1-bit input data at user $k$.

Let $F_{k}$ denote the maximum computation capacity of user $k$, which is measured by the number of CPU cycles per second.
Denoting $T$ as the maximal latency of all users, we can obtain the following local computation latency constraints
\begin{equation}\label{sys1eq1_2}
(R_{k}-d_{k}) C_{k}\leq F_{k} T, \quad \forall k\in\mathcal K,
\end{equation}
which can be equivalent to
\begin{equation}\label{sys1eq1_22}
d_{k} \geq \frac{R_{k} C_{k}-F_{k} T}
{C_{k}}, \quad \forall k\in\mathcal K.
\end{equation}
Defining $f_{k}$ as the required CPU frequency for computing each CPU cycle at user $k$, it needs to satisfy
\begin{align}\label{sys4}
f_k=\frac{\mathcal C_{k}^{\text{Loc}}}{T}.
\end{align}
The local energy consumption can be expressed as \cite{NOMAmecWang}
\begin{align}\label{eloc}
E_{k}^{\text{Loc}}=\alpha f^2_k\mathcal C_{k}^{\text{Loc}},
\end{align}
where $\alpha$ represents the effective capacitance coefficient which is determined by the chip architecture.
By substituting (\ref{sys1eq1}) and (\ref{sys4}) into (\ref{eloc}), we have
\begin{align}
E_{k}^{\text{Loc}}=\alpha\frac{(R_{k}-d_{k})^3 C^3_{k}}{T^2}.
\end{align}
It is also assumed that the edge cloud has finite computation capacity $F$.
As a result, the offloading data of all users should satisfy the following computation constraint:
\begin{equation}\label{sys1eq1_3}
\sum_{k=1}^K d_{k} C_{k}\leq F.
\end{equation}

\subsection{Offloading Model}
Denote the bandwidth of the network by $B$, the power of the additive white Gaussian noise by $\sigma^2$ and the maximum transmit power of user $k$ by $P_k$.
Denote the transmit signal of user $k$ by $\sqrt{p_k}s_k$, where $p_k$ is the transmit power, which satisfies
\begin{align}
0\leq p_k\leq P_k,
\end{align}
and $s_k$ is the signal with zero mean and unit variance.
With the help of RIS, the received signal at the BS is
{\begin{equation}\label{sys2eq2}
\pmb y=
\sum_{k=1}^K \left(\pmb h_{D,k}^H+\pmb h_{k}^H \pmb \Theta \pmb G\right)  \sqrt{p_k}s_k+\pmb n_k^H ,
\end{equation}
where $\pmb h_{D,k}^H\in\mathbb C^{M}$ denotes the direct link channel response from user $k$ to the BS,} 
$\boldsymbol G\in\mathbb C^{N \times M}$ and $\boldsymbol h_{k}\in\mathbb C^{N}$ respectively denote the channel responses from BS to RIS and from RIS to user $k$,  
the phase shift matrix of the RIS is $\pmb \Theta =\text{diag} (e^{i\theta_{1}}, \cdots, e^{i\theta_{N}})\in\mathbb C^{N\times  N}$  with $\theta_{n}\in[0,2\pi]$, $n\in\mathcal N=\{1,\cdots,N\}$,
and $\pmb n_k\in\mathbb C^{M}$ denotes the additive white Gaussian
noise at the BS with $\pmb n_k \sim \mathcal C \mathcal N(\pmb 0,  \sigma^2\bf{I} )$.

By allowing the BS to properly design the decoding order of NOMA and employing time-sharing among different decoding orders,
the capacity region for the $K$ users is achievable and it corresponds to the following polymatroid \cite{tse2005fundamentals}
\begin{align}\label{sys1eq6_0}
\mathcal C(\pmb p,  \pmb \theta)=&
\bigg\{
\pmb r \in \mathbb R^{N}:
\sum_{k\in\mathcal K'}r_{k}   \\\nonumber \leq
&B \log_2 \left( 1+ {\sum_{k\in\mathcal K'}\frac{{\pmb h_{D,k}^H+}\pmb h_{k}^H \pmb \Theta \pmb G}{\sigma^2}  {p_k}} \right),
\forall \mathcal K'\subseteq \mathcal K \setminus \emptyset
\bigg\},
\end{align}
where $r_k$ denotes the achievable rate of user $k$,
$\pmb r =[r_1,\cdots, r_K]$,
$\pmb\theta=[\theta_{1},\cdots,\theta_{N}]$,
$\emptyset$ is an empty set, and $\mathcal K'\subseteq \mathcal K\setminus \emptyset$ means that $\mathcal K'$ is a non-empty subset of $\mathcal K$.

Denoting the uplink transmission time by $t$, the energy consumption for offloading at user $k$ is given by
\begin{eqnarray}\label{sys1eq6}
E_{k}^{\text{Off}}= tp_k, \quad \forall k\in\mathcal K.
\end{eqnarray}
To satisfy that $d_k$ bits can be offloaded to the BS, we have
\begin{eqnarray}\label{sys1eq62}
tr_k\geq d_k, \quad \forall k\in\mathcal K.
\end{eqnarray}

\subsection{Problem Formulation}
{In light of some effectively channel estimation methods for RIS-aided wireless networks were proposed in existing works \cite{yuan2020reconfigurable,channelestimationyi,channelestimationzheng, channelestimationwang, wei2020channel}. Then, in this paper, we have the assumption that the CSI is perfectly obtained at the transmitter side.}

 Based on the above model, the sum user energy minimization problem is formulated as:
\begin{subequations}\label{sys1min1}
	\begin{align}
	\mathop{\min}_{\pmb d,t,\pmb p,\pmb r,\pmb \theta}\quad
	&\sum_{k=1}^K (\alpha\frac{(R_{k}-d_{k})^3 C^3_{k}}{T^2}+tp_k)
	\\
	\textrm{s.t.}\quad
	& \sum_{k=1}^K d_{k} C_{k}\leq F,\\
	& tr_k\geq d_k, \quad \forall k\in\mathcal K,\\
	& \pmb r \in \mathcal C(\pmb p,  \pmb \theta),\\
	& 0\leq t \leq T,\\
	&  D_{k} \leq d_{k} \leq R_{k}, \quad \forall k\in\mathcal K,\\
	&  0\leq p_k\leq P_k, \quad \forall k\in\mathcal K,
	\end{align}
\end{subequations}
where $\pmb d=[d_{1},\cdots,d_{K}]$,  $\pmb p=[p_1, \cdots, p_K]$ and $D_{k}=\max\left\{\frac{R_{k} C_{k}-F_{k} T}
{C_{k}},0\right\}$ is obtained from equation (\ref{sys1eq1_22}).
The objective function (\ref{sys1min1}a) represents the total energy consumption of all users including both offloading energy and computing energy. Constraint (13b) shows the maximum computation ability at the BS, while  each user can successfully transmit the task bits as shown in (13c). The rate region of NOMA is expressed in constraint (13d).
Constraint (\ref{sys1min1}e) ensures that the offloading can be completed in time requirement of $T$ for all users. Constraint (13f) shows the minimum and maximum requirements of the offloaded bits of each user. The maximum power constraint is given in (13g).

\section{An Efficient algorithm to Solve Problem (\ref{sys1min1})}

Obviously, Problem (\ref{sys1min1}) is nonconvex due to the nonconvex constraints (\ref{sys1min1}c) and (\ref{sys1min1}d). In order to solve Problem (\ref{sys1min1}), we propose an efficient iterative algorithm based on the alternating method. In particular, for given phase shift vector $\pmb \theta$, we use the dual method to tackle this subproblem and obtain the optimal offloading bits $\pmb d^*$, transmission time $t^*$, power allocation $\pmb p^*$ and use a time sharing technique to obtain the optimal decoding order and transmission rate $\pmb r^*$.
Then for given $\pmb p$, we use the penalty method to optimize $\pmb d$, $t$, $\pmb r$ and $\pmb \theta$.
The sub-optimal solution to Problem (\ref{sys1min1}) is obtained via iteratively solving the aforementioned two subproblems.

In the following, we show how to solve one subproblem of Problem (\ref{sys1min1}) with given $\pmb \theta$ and use the penalty method to solve another subproblem with given $\pmb p$.

\subsection{Solve Problem (\ref{sys1min1}) for given $\pmb \theta$}

Firstly, for given $\pmb \theta$, the optimization problem (\ref{sys1min1}) can be reformulated as

\begin{subequations}\label{sys1min2}
	\begin{align}
	\mathop{\min}_{\pmb d,t,\pmb p,\pmb r}\quad
	&\sum_{k=1}^K \left(\alpha\frac{(R_{k}-d_{k})^3 C^3_{k}}{T^2}+tp_k\right)
	\\
	\textrm{s.t.}\quad
	& \sum_{k=1}^K d_{k} C_{k}\leq F,\\
	& tr_k\geq d_k, \quad \forall k\in\mathcal K,\\
	& \pmb r \in \mathcal C(\pmb p),\\
	& 0\leq t \leq T,\\
	&  D_{k} \leq d_{k} \leq R_{k}, \quad \forall k\in\mathcal K,\\
	&  0\leq p_k\leq P_k, \quad \forall k\in\mathcal K.
	\end{align}
\end{subequations}

\begin{theorem}
	When $t>0$, the optimal transmission time $t$ of Problem (\ref{sys1min2}) is $t=T$.\footnote{Compared to \cite{EE}, the optimization problem is formulated in a different manner and the method in the proof is different.}
\end{theorem}

\pmb{Proof:}
For given $\pmb d$, Problem (\ref{sys1min2}) can be written as
\begin{subequations}\label{sys1min22}
	\begin{align}
	\mathop{\min}_{t,\pmb p,\pmb r}\quad
	&t\sum_{k=1}^K p_k
	\\
	\textrm{s.t.}\quad
	& tr_k\geq d_k, \quad \forall k\in\mathcal K,\\
	& \pmb r \in \mathcal C(\pmb p),\\
	& 0\leq t \leq T,\\
	&  0\leq p_k\leq P_k, \quad \forall k\in\mathcal K.
	\end{align}
Assume that $(t^*,p^*_k)$ is the optimal solution of Problem (\ref{sys1min22}) with $0<t^*<T$.
In order to minimize the objective function of Problem (\ref{sys1min22}), we have
\begin{align}
r^*_k=\frac{d_k}{t^*}.
\end{align}
Thus constraint (\ref{sys1min22}c) can be rewritten as
{\begin{align}\label{sys1eq66}
	&\mathcal C(\pmb p,  \pmb \theta)=
	\bigg\{
	\pmb r \in \mathbb R^{N}:
	\sum_{k\in\mathcal K'}d_k \\\nonumber \leq
	&t^*B \log_2 \left( 1+ {\sum_{k\in\mathcal K'}\frac{ \left\|\pmb h_{D,k}^H+\pmb h_{k}^H \pmb \Theta \pmb G\right\|^2}{\sigma^2}  {p^*_k}} \right),
	\forall \mathcal K'\subseteq \mathcal K \setminus \emptyset
	\bigg\}.
\end{align}}
\end{subequations}
Define $s=T/t^*$, thus $s>1$.
According to the monotonicity of (\ref{sys1eq66}), we have
{\begin{align}\label{s16}
\sum_{k\in\mathcal K'}d_k &\leq t^*B \log_2 \left( 1+ {\sum_{k\in\mathcal K'}\frac{\left\|\pmb h_{D,k}^H+\pmb h_{k}^H \pmb \Theta \pmb G\right\|^2}{\sigma^2}  {p^*_k}} \right)\\\nonumber
&< st^*B \log_2 \left( 1+ \frac{1}{s}{\sum_{k\in\mathcal K'}\frac{\left\|\pmb h_{D,k}^H+\pmb h_{k}^H \pmb \Theta \pmb G\right\|^2}{\sigma^2}  {p^*_k}} \right),
\end{align}}
where the last inequality in (\ref{s16}) follows from the fact that function $x \log_2(1+1/x)$ is monotonically increasing when $x>0$,
and we can find a coefficient $s'$ ($0<s'<1$) that satisfies the following condition
{\begin{align}
&\sum_{k\in\mathcal K'}d_k \leq t^*B \log_2 \left( 1+ {\sum_{k\in\mathcal K'}\frac{\left\|\pmb h_{D,k}^H+\pmb h_{k}^H \pmb \Theta \pmb G\right\|^2}{\sigma^2}  {p^*_k}} \right)\\\nonumber
&= st^*B \log_2 \left( 1+ \frac{1}{s}{\sum_{k\in\mathcal K'}\frac{\left\|\pmb h_{D,k}^H+\pmb h_{k}^H \pmb \Theta \pmb G\right\|^2}{\sigma^2}  s'{p^*_k}} \right).
\end{align}}
According to the objective function of (\ref{sys1min22}), we have
\begin{align}
\sum_{k=1}^K st^*\times s'\frac{1}{s}p^*_k=\sum_{k=1}^K s'T\times \frac{t^*p^*_k}{T}<\sum_{k=1}^K t^*p^*_k,
\end{align}
which means when $0<t^*<T$, we can always construct a new set of solution $t'=T$ and $p'_k=s'\frac{t^*p^*_k}{T}$ that makes the objective function of Problem (\ref{sys1min22}) smaller than when the solutions are $t^*$ and $p^*_k$.
Therefore, when $t>0$, the optimal solution of $t$ is $t=T$.

It is found that all constraints in Problem (\ref{sys1min2}) are convex except (\ref{sys1min2}c).
Constraint (\ref{sys1min2}c) is neither convex nor concave because its Hessian is indefinite which has a positive and a negative eigenvalue. However, function (\ref{sys1min2}c) is quasiconcave for all $t\geq 0, d_k \geq 0$, since the suplevel sets $\{t, r_k \in \pmb R_+| tr_k\geq d_k \}$ are convex sets for all $d_k$\cite{boyd2004convex}.

Therefore, Problem ($\ref{sys1min2}$) is convex and satisfies the Slater's condition, based on which the strong duality between Problem ($\ref{sys1min2}$) and its dual problem holds. The partial Lagrangian function of Problem ($\ref{sys1min2}$) is
\begin{align}\label{lagarangian}
&\mathcal L(\pmb d,t,\pmb p,\pmb r,\lambda,\pmb \mu)= \sum_{k=1}^K (\alpha\frac{(R_{k}-d_{k})^3 C^3_{k}}{T^2}+tp_k)\\ \nonumber
&+\lambda( \sum_{k=1}^K d_kC_k-F)+\sum_{k=1}^K\mu_k(d_k-tr_k),
\end{align}
where $\lambda$ and $\mu_k$ are the Lagrange dual variables.
The Lagrange dual function is given by
\begin{subequations}\label{sys1min3}
	\begin{align}
	g(\lambda,\pmb \mu)=\mathop{\min}_{\pmb d, t,\pmb p,\pmb r} \mathcal L&(\pmb d,t,\pmb p,\pmb r,\lambda,\pmb \mu)\\
	s.t. \quad &\pmb r \in \mathcal C(\pmb p),\\
	&  0\leq t \leq T,\\
	&  D_{k} \leq d_{k} \leq R_{k}, \quad \forall k\in\mathcal K,\\
	&  0\leq p_k\leq P_k, \quad \forall k\in\mathcal K.
	\end{align}
\end{subequations}
Accordingly, the Lagrange dual problem can be expressed as

\begin{subequations}\label{sys1min4}
	\begin{align}
	\mathop{\max}_{\lambda,\pmb \mu}& \quad g(\lambda,\pmb \mu)\\
	s.t.
	&\quad  \mu_k \geq 0, \quad \forall k\in\mathcal K,\\
	&\quad  \lambda \geq 0.
	\end{align}
\end{subequations}
To tackle Problem (\ref{sys1min2}), we solve Problem (\ref{sys1min3}) to obtain $g(\lambda,\pmb \mu)$ with given $\lambda,~\pmb \mu$ and then solve Problem (\ref{sys1min4}) to update the Lagrange variables $\lambda$ and $\pmb \mu$. The optimal solutions $\pmb d^*,~t^*,~\pmb p^*,~\pmb r^*$ can be obtained by iteratively solving these two problems.

\subsubsection{Evaluating $g(\lambda,\pmb\mu)$ by solving Problem (\ref{sys1min3})}
First, for given $\lambda,~\pmb\mu$, we obtain the dual function $g(\lambda,\pmb\mu)$ by solving Problem (\ref{sys1min3}). The Lagrangian in (\ref{lagarangian}) can be rewritten as follows

\begin{align}\label{lagarangian2}
&\mathcal L(\pmb d,t,\pmb p,\pmb r, \lambda,\pmb \mu)=
\sum_{k=1}^K (\lambda d_kC_k+\alpha\frac{\left(R_{k}-d_{k}\right)^3 C^3_{k}}{T^2}+\mu_kd_k)\\\nonumber
&+\sum_{k=1}^K (tp_k-\mu_ktr_k)-\lambda F.
\end{align}

Note that the objective function as well as all the constraints of Problem (\ref{sys1min3}) is decoupled for $\pmb d$ and other variables. Then, Problem (\ref{sys1min3}) can be decomposed into $2K$ subproblems through dropping the constant term $\lambda F$ as follows,

\begin{subequations}\label{p1}
	\begin{align}
	\mathop{\min}_{\pmb d} \quad &\sum_{k=1}^K (\lambda d_kC_k+\frac{\alpha \left(R_{k}-d_{k}\right)^3C^3_k}{T^2}+\mu_kd_k)\\
	s.t. \quad  & D_k \leq d_k \leq R_k.
	\end{align}
\end{subequations}
and
\begin{subequations}\label{p2}
	\begin{align}
	\mathop{\min}_{\pmb r,\pmb p, t}\quad  &\sum_{k=1}^K (tp_k-\mu_ktr_k)\\
	s.t. \quad &\pmb r \in \mathcal C(\pmb p),\\
	& 0 \leq t \leq T,\\
	&  0\leq p_k\leq P_k, \quad \forall k\in\mathcal K.
	\end{align}
\end{subequations}

%
%

It is obvious to find that Problem (\ref{p1}) is a convex problem for $d_k$, and the Slater conditions hold. The following theorem gives the optimal solution of Problem (\ref{p1}).
\begin{theorem}
	The optimal solution of Problem (\ref{p1}) is
	\begin{align}
	d^*_k=\max \{R_k-\sqrt{\frac{\lambda C_k+\mu_k}{3\alpha C_k}}\frac{T}{C_k}, D_k\}, \forall k\in \mathcal K.
	\end{align}
\end{theorem}

\pmb{Proof:}
	Since each user $k$ is independent of each other, the Lagrangian function corresponding to user $k$ is
	\begin{align}\label{p111}
	\mathcal L_k=& \lambda d_kC_k+\frac{\alpha \left(R_{k}-d_{k}\right)^3C^3_k}{T^2}+\mu_kd_k-\bar \beta_k(d_k-D_k)\\\nonumber-&\underline\beta_k(R_k-d_k),
	\end{align}
	where $\bar \beta_k$, $\underline\beta_k$ are the Lagrangian dual multipliers associated with constraint (19b). The Karush-Kuhn-Tucke (KKT) conditions of Problem (\ref{p1}) are as follows
	
	\begin{subequations}\label{p2222}
		\begin{align}
		& \frac{\partial \mathcal L_k }{\partial d_k} =\lambda C_k+\mu_k-\frac{3\alpha \left(R_{k}-d_{k}\right)^2C^3_k}{T^2}-\bar \beta_k+\underline\beta_k=0, \\
		& \bar \beta^*_k(d^*_k-D_k)=0,\\ &\underline\beta^*_k(R_k-d^*_k)=0,\\
		&  D_k \leq d^*_k \leq R_k,\\
		& \bar \beta^*_k\geq 0,\underline\beta^*_k \geq 0.
		\end{align}
	\end{subequations}
	Substituting (\ref{p2222}d) and (\ref{p2222}e) into (\ref{p2222}b) and (\ref{p2222}c), we have $\bar \beta^*_k=\underline\beta^*_k=0$ when $D_k<d^*_k<R_k$.
	Solving conditions (\ref{p2222}a), we can obtain the optimal $d^*_k$, which is given by
	\begin{align}
	d^*_k=\left.\left(R_k-\sqrt{\frac{\lambda C_k+\mu_k}{3\alpha C_k}}\frac{T}{C_k}\right)\right|^{R_k}_{D_k},
	\end{align}
	where $\left.x\right|^a_b=\min\{\max\{x,b\},a\}$.
	Since
	\begin{align}
	R_k-\sqrt{\frac{\lambda C_k+\mu_k}{3\alpha C_k}}\frac{T}{C_k}\leq R_k,
	\end{align}
	we have
	\begin{align}
	d^*_k=\max \{R_k-\sqrt{\frac{\lambda C_k+\mu_k}{3\alpha C_k}}\frac{T}{C_k}, D_k\}.
	\end{align}
	Thus, Theorem 2 is proved.
 

 The objective function of Problem (\ref{p2}) is a linear function of $t$. If $\sum_{k=1}^K (p_k-\mu_kr_k)> 0$, then the optimal transmission time is $t^*=0$, which means that all tasks need to be computed locally, and we can set $p^*_k=r^*_k=0$. If $\sum_{k=1}^K (p_k-\mu_kr_k)\leq 0$, then the optimal transmission time is $t^*=T$ and in this case Problem (\ref{p2}) can be transformed into
\begin{subequations}\label{p3}
	\begin{align}
	\mathop{\min}_{\pmb r,\pmb p}\quad &T\sum_{k=1}^K\left(p_k-\mu_kr_k\right)\\
	s.t. \quad &\pmb r \in \mathcal C(\pmb p),\\
	&0\leq p_k \leq P_k, \forall k \in \mathcal K.
	\end{align}
\end{subequations}
{To solve Problem (\ref{p3}), we first consider the following subproblem with given $\pmb p$
\begin{subequations}\label{p44}
	\begin{align}
		\mathop{\max}_{\pmb r}\quad &\sum_{k=1}^K\mu_kr_k\\
		s.t. \quad &\pmb r \in \mathcal C(\pmb p),
	\end{align}
\end{subequations}
whose optimal solution can be obtained via the following theorem as shown in \cite{leD}.}

\begin{theorem}
{	The optimal rate vector of Problem (\ref{p44})
	is given by one vertex $\pmb r^*_{\pi}=\left[r^*_{\pi(1)},r^*_{\pi(2)},\dots,r^*_{\pi(K)}\right]$ of polymatroid $\mathcal C(\pmb p, \pmb \theta)$, where $\pmb r^*_{\pi}$ represents the optimal rate vector under the optimal decoding order $\pmb\pi=[\pi(1),\dots,\pi(K)]$ determined by $\mu_{\pi(1)}\geq\dots\geq\mu_{\pi(K)}\geq 0$, and $r^*_{\pi(k)}$ is the achievable rate of user $\pi(k)$ given by \cite{tse2005fundamentals}}
	{\begin{align}\label{rates}
			r^*_{\pi(k)}=B\log_2\left(\frac{1+ \sum_{i=1}^k {\|\pmb h_{D,\pi(i)}^{H}+\pmb h_{\pi(i)}^{H} \pmb \Theta \pmb G}\|^2 {p_{\pi(i)}}}{1+ \sum_{i=1}^{k-1} { \|\pmb h_{D,\pi(i)}^{H}+\pmb h_{\pi(i)}^{H} \pmb \Theta \pmb G}\|^2 {p_{\pi(i)}}}\right).
	\end{align}}
\end{theorem}	

{According to Theorem 3, the optimal decoding order is determined by the order of the Lagrange multipliers. It has been proven in \cite[Lemma 3.2]{leD} that the complexity of solving problem (\ref{p44}) is $\mathcal{O}(K\log_2(K))$.
Using (\ref{rates}), Problem (\ref{p3}) can be rewritten as
\begin{subequations}\label{p5}
	\begin{align}
	\mathop{\min}_{\pmb p}\quad & T \sum_{k=1}^K\left(p_k-\left(\mu_{\pi(k)}-\mu_{\pi(k+1)}\right)
	\right.\\
	\nonumber
	~~~&
	\left.
	\times B \log_2\left(1+ {\sum_{i=1}^k \left\|\pmb h_{D,\pi(i)}^{H}+\pmb h_{\pi(i)}^{H} \pmb \Theta \pmb G\right\|^2  {p_{\pi(i)}}}\right)\right)\\
	s.t. ~~0&\leq p_k \leq P_k, \forall k \in \mathcal K.
	\end{align}
\end{subequations}}
For the convenience of notation, we set $\mu_{\pi(K+1)}=0$. Note that Problem (\ref {p5}) is convex with respect to $\pmb p$, and it can be easily solved by the standard toolbox, such as CVX \cite{cvx}. Let $\pmb p^*$ denote the optimal solution to Problem (\ref{p5}) and the optimal $\pmb r^*$ can be obtained by substituting $\pmb p^*$ into equation (\ref{rates}).

\subsubsection{Using time sharing method to obtain primal optimal $\pmb r^*$ and decoding order for Problem (\ref{sys1min2})}
It is worth mentioning that the decoding order in Problem (\ref{sys1min2}) is determined by $\pmb \mu$. However, if there exists $k$ and $k'$ such that $\mu_k=\mu_k'$, the decoding order as well as the transmission rate $\pmb r^*$ is not unique. In this case, the primal optimal solution can be obtained by time-sharing method.

More specifically, define $\pmb {\chi}_1,\cdots,\pmb {\chi}_L $ as $L$ disjoint subsets in which $\mu_k=\mu_k'~(\forall k, k'\in \pmb\chi_l) $ are equal. Define $|\pmb \chi_l|$ as the cardinality of the set $\pmb \chi_l$, and there are $|\pmb \chi_l|\geq 2$, $\forall l=1,\dots, L$. To find the optimal decoding order by time-sharing, we can divide the transmission time $t$ into $|\Psi|$ time slots, where $\Psi$ can be expressed as
\begin{align}
\Psi=\{1,...,\prod \nolimits_{l=1}^L |\pmb \chi_l|\}.
\end{align}
Assuming $r_k^\varphi$ is the achievable rate in time slot $\varphi$, and each length of time slots is obtained by solving the following problem

\begin{subequations}\label{time}
	\begin{align}
	\text{Find}\quad
	&\tau^1,\dots,\tau^{|\Psi|}
	\\
	\textrm{s.t.}\quad
	& \sum_{\varphi=1}^{|\Psi|} \tau^\varphi r_k^\varphi\geq d_k, \quad \forall k\in\mathcal K,\\
	& \sum_{\varphi=1}^{|\Psi|} \tau^\varphi \leq t^*,\\
	&  0 \leq \tau^\varphi\leq t^*, \quad \forall k\in\mathcal K.
	\end{align}
\end{subequations}
Problem (\ref{time}) is a linear problem and can be easily solved by CVX.

\subsubsection{Obtaining the optimal $\lambda^*,\pmb\mu^*$ by solving Problem (\ref{sys1min4})}

Since $g(\pmb \lambda, \pmb \mu)$ is obtained, $\lambda^*,\pmb\mu^*$ can be obtained by solving Problem (\ref{sys1min4}). Although $g(\pmb \lambda, \pmb \mu)$ is convex with respect to $\lambda, \pmb \mu$, yet may not be differentiable in general. Therefore, $\lambda^*,\pmb\mu^*$ can be updated as follows

\begin{align}\label{lagrange1}
\lambda(x+1)=\left[\lambda(x)+\delta_1(x)F-\sum_{k=1}^K d^*_kC_k  \right]^+,
\end{align}
\begin{align}\label{lagrange2}
\mu_k(x+1)=\left[\mu_k(x)+\delta_2(x)t^*r^*_k- d^*_k  \right]^+,
\end{align}
where $[a]^+$ represents the function $\max\{a,0\}$, $x$ is the iteration number, and $\delta_1(x), \delta_2(x)$ are the dynamically chosen step size sequences which are based on suitable estimates.

\subsection{Solve Problem (\ref{sys1min1}) for given $\pmb p$}
For given $\pmb p$, Problem (\ref{sys1min1}) can be rewritten as follows,

\begin{subequations}\label{sys1minp}
	\begin{align}
	\mathop{\min}_{\pmb d,  t, \pmb r, \pmb \theta}\quad
	&\sum_{k=1}^K (\alpha\frac{(R_{k}-d_{k})^3 C^3_{k}}{T^2}+tp_k)
	\\
	\textrm{s.t.}\quad
	& \sum_{k=1}^K d_{k} C_{k}\leq F,\\
	& tr_k\geq d_k, \quad \forall k\in\mathcal K,\\
	& \pmb r \in \mathcal C(\pmb \theta),\\
	& 0\leq t \leq T,\\
	&  D_{k} \leq d_{k} \leq R_{k}, \quad \forall k\in\mathcal K.
	\end{align}
\end{subequations}
According to Theorem 1, we have $t^*=0$ or $t^*=T$.
When $t^*=0$, which means that all tasks should be computed locally and the optimal solution can be $\pmb d^*=\pmb 0, \pmb r^*=\pmb 0, \pmb \theta^*=\pmb 0$.
When $t^*=T$, Problem (\ref{sys1minp}) can be rewritten as
\begin{subequations}\label{sys1minp2}
	\begin{align}
	\mathop{\min}_{\pmb d, \pmb r, \pmb \theta}\quad
	&\sum_{k=1}^K \alpha\frac{(R_{k}-d_{k})^3 C^3_{k}}{T^2}
	\\
	\textrm{s.t.}\quad
	& \sum_{k=1}^K d_{k} C_{k}\leq F,\\
	& d_k\leq Tr_k, \quad \forall k\in\mathcal K,\\
	& \pmb r \in \mathcal C(\pmb \theta),\\
	&  D_{k} \leq d_{k} \leq R_{k}, \quad \forall k\in\mathcal K.
	\end{align}
\end{subequations}
It is not difficult to find that all constraints in Problem (\ref{sys1minp2}) are convex except (\ref{sys1minp2}d).
%
Define $\pmb u=[u_1, \cdots, u_N]^H$, where $u_n=e^{i\theta_{n}}, n=1,\dots,N$ and we have
{\begin{align}
&\left\|\pmb h^H_{D,k}+\pmb h_{k}^H \pmb \Theta \pmb G\right\|^2\\\nonumber
&= \pmb h^H_{D,k}\pmb h_{D,k}+\pmb h_{k}^H \pmb \Theta \pmb G\pmb h_{D,k}+\pmb h_{D,k}^H\pmb G^H\pmb \Theta^H\pmb h_{k}\\\nonumber&~~+\pmb h_{k}^H \pmb \Theta \pmb G\pmb G^H\pmb \Theta^H\pmb h_{k}\\\nonumber
&=\pmb h^H_{D,k}\pmb h_{D,k}+\pmb u^H\pmb V_k\pmb h_{D,k}+\pmb h^H_{D,k}\pmb V_k^H\pmb u+\pmb u^H\pmb V_k\pmb V_k^H\pmb u,
\end{align}
where $\pmb V_k=\text{diag}(\pmb h_{k}^H)\pmb G \in \mathbb C^{N\times M} $.}
We introduce a slack vector $\pmb \eta=[\eta_1,\eta_2,\dots,\eta_K]$, which satisfies
\begin{align}\label{sys1eq1112}
{\eta_k\leq\pmb h^H_{D,k}\pmb h_{D,k}+\pmb u^H\pmb V_k\pmb h_{D,k}+\pmb h^H_{D,k}\pmb V_k^H\pmb u+\pmb u^H\pmb V_k\pmb V_k^H\pmb u.}
\end{align}
Constraint (\ref{sys1eq1112}) is non-convex with respect to $\pmb\eta$ and $\pmb u$, and we use the successive convex approximation (SCA) method to approximate it by a convex constraint. Let
$\pmb u^{(r)}$ denote the vector $\pmb u$ in the $r$-th iteration of
the SCA method. Constraint (\ref{sys1eq1112}) can be approximated by
{\begin{align}\label{sys1eq1113}
\eta_k\leq& 2\text{Re}\left(\pmb u^{(r)H}\pmb V_k\left(\pmb u-\pmb u^{(r)}\right)\right)+\pmb u^{(r)H}\pmb V_k\pmb V_k^H\pmb u^{(r)}\\\nonumber
&+\pmb h^H_{D,k}\pmb h_{D,k}+\pmb u^H\pmb V_k\pmb h_{D,k}+\pmb h^H_{D,k}\pmb V_k^H\pmb u.
\end{align}}
Then, Problem (\ref{sys1minp2}) can be reformulated as
\begin{subequations}\label{sys1minp3}
	\begin{align}
	\mathop{\min}_{\pmb d, \pmb r, \pmb u, \pmb \eta}\quad
	&\sum_{k=1}^K \alpha\frac{(R_{k}-d_{k})^3 C^3_{k}}{T^2}
	\\
	\textrm{s.t.}\quad
	& \sum_{k=1}^K d_{k} C_{k}\leq F,\\
	&  D_{k} \leq d_{k} \leq R_{k}, \quad \forall k\in\mathcal K,\\
	& d_k\leq Tr_k, \quad \forall k\in\mathcal K,\\
	&\sum_{k\in\mathcal K'}r_{k} \leq
     B \log_2 \left( 1+ {\sum_{k\in\mathcal K'}\frac{\eta_k}{\sigma^2}  {p_k}} \right),
     \forall \mathcal K'\subseteq \mathcal K \setminus \emptyset,\\
    & |u_n|= 1, \quad \forall n\in\mathcal N,\\
	&(\ref{sys1eq1113}), \quad \forall k\in\mathcal K.
	\end{align}
\end{subequations}
Noting that (\ref{sys1minp3}f) is still a nonconvex constraint, we consider to utilize the penalty method and Problem (\ref{sys1minp3}) can be formulated as
\begin{subequations}\label{sys1minp44}
	\begin{align}
	\mathop{\min}_{\pmb d, \pmb r, \pmb u, \pmb \eta}\quad
	&\sum_{k=1}^K \alpha\frac{(R_{k}-d_{k})^3 C^3_{k}}{T^2}+Q\sum_{n\in\mathcal N}\left(1-|u_n|^2\right)
	\\
	\textrm{s.t.}\quad
	& \sum_{k=1}^K d_{k} C_{k}\leq F,\\
	&  D_{k} \leq d_{k} \leq R_{k}, \quad \forall k\in\mathcal K,\\
	& d_k\leq Tr_k, \quad \forall k\in\mathcal K,\\
    &(\ref{sys1minp3}e),\\
	& |u_n|\leq 1, \quad \forall n\in\mathcal N,\\
	&(\ref{sys1eq1113}), \quad \forall k\in\mathcal K,
	\end{align}
\end{subequations}
where $Q$ is a large positive coefficient. Obviously, $Q\sum_{n\in\mathcal N}\left(1-|u_n|^2\right)$ enforces that when $1-|u_n|^2=0, \forall n\in\mathcal N$, $\pmb u$ obtains the optimal solution.
To handle the nonconvexity of  Problem (\ref{sys1minp44}), SCA is utilized to approximate the objective function as
\begin{align}
\sum_{k=1}^K \alpha\frac{(R_{k}-d_{k})^3 C^3_{k}}{T^2}&+Q\sum_{n\in\mathcal N}\left(1-|u^{r}_n|^2\right)\\\nonumber
&-2Q\sum_{n\in\mathcal N}\mathbb{R}\left(u^{r}_n\left(u_n- u^{r}_n\right)\right),
\end{align}
where $ u_n^{(r)}$ denotes $u_n$ in the $r$-th iteration of the SCA method.

Based on the aforementioned approximations, by dropping the constant term $Q\sum_{n\in\mathcal N}\left(1-| u^{r}_n|^2\right)$, Problem (\ref{sys1minp}) can be rewritten as
\begin{subequations}\label{sys1mindp44}
	\begin{align}
	\mathop{\min}_{t, \pmb r, \pmb u, \pmb \eta}\quad
	&\sum_{k=1}^K \alpha\frac{(R_{k}-d_{k})^3 C^3_{k}}{T^2}-2Q\sum_{n\in\mathcal N}\mathbb{R}\left( u^{r}_n\left(u_n- u^{r}_n\right)\right)\\
	\textrm{s.t.}\quad
	& d_k\leq Tr_k, \quad \forall k\in\mathcal K,\\
	& 0\leq t \leq T,\\
	& |u_n|\leq 1, \quad \forall n\in\mathcal N,\\
	&(\ref{sys1minp3}e),\\
	&(\ref{sys1eq1113}), \quad \forall k\in\mathcal K.
	\end{align}
\end{subequations}
Since Problem (\ref{sys1mindp44}) is a convex problem, we can utilize the interior point method or CVX\cite{cvx} to obtain the optimal solution of Problem (\ref{sys1mindp44}). Moreover, the solution of $\pmb\theta$ can be updated based on $\pmb u$.

\subsection{Iterative Algorithm for solving Problem (\ref{sys1min1})}
The alternating procedure for solving Problem (\ref{sys1min1}) is given in Algorithm 1. We utilize a block coordinate descent method to alternately solve Problem (\ref{sys1min2}) and Problem (\ref{sys1minp}), and finally obtain the suboptimal solution.
Since the objective value decreases with the iteration number and has a finite lower bound, the algorithm is guaranteed to converge.

The procedure for solving Problem (\ref{sys1min2}) is shown as Algorithm 2. The dual method is introduced to find the optimal closed-form solutions of Problem (\ref{sys1min2}) as well as the optimal decoding order. To solve Problem (\ref{sys1minp}), the penalty method is utilized to approximately transform this problem to Problem (\ref{sys1mindp44}) and the procedure of solving Problem (\ref{sys1mindp44}) is shown in Algorithm 3.

\begin{algorithm}[h]
	\caption{:Block Coordinate Descent Algorithm for Problem (\ref{sys1min1})}
	\label{alg:Framwork1}
	\begin{algorithmic}[1]
		\STATE \pmb{Initialize} $\pmb d^{(0)}$,~ $t^{(0)}$,~ $\pmb p^{(0)}$,~ $\pmb r^{(0)}$,~ $\pmb \theta^{(0)}$, the tolerance $\tau$, iteration variable $a=0$.
		\REPEAT
		\STATE With $\pmb \theta^{(a)}$, update $\pmb d^{(a)},~t^{(a)},~ \pmb r^{(a)},~ \pmb r^{(a)}$ by solving Problem (\ref{sys1min2}).
		\STATE With $\pmb p^{(a)}$, update $\pmb d^{(a)}$, $t^{(a)}$, $\pmb r^{(a)}$, $\pmb \theta^{(a)}$ by solving Problem (\ref{sys1minp}).
        \STATE Set $a=a+1$.
		\UNTIL the objective value of Problem (\ref{sys1min1}) converges.
	\end{algorithmic}
\end{algorithm}
\begin{algorithm}[h]
	\caption{:Alternating Procedure for Solving Problem (\ref{sys1min2})}
	\label{alg:Framwork2}
	\begin{algorithmic}[1]
   \STATE \pmb{Initialize} $\pmb \theta^{(a)}$, the tolerance $\tau$, iteration variable $i=0$.
   \REPEAT
   \STATE Update Lagrange variables $\lambda, \pmb \mu$ using Eq. (\ref{lagrange1}) and Eq. (\ref{lagrange2}).
   \STATE Set the permutation $\pmb\pi$, s.t. $\mu_{\pi(1)}\geq\dots\geq\mu_{\pi(K)}\geq 0$.
   \STATE Update $\pmb p$ by solving Problem (\ref{p5}).
   \STATE Update $\pmb r$ by using Eq. (\ref{rates}).
   \STATE Update $\pmb d$ and $t$ by solving Problem (\ref{p1}) and (\ref{p2}).
   \STATE Calculate the objective function of Problem (\ref{sys1min1}) and the value is denoted by $L(i)$, $i=i+1$.
   \UNTIL $L(i)-L(i-1)<\tau$.
   \STATE \pmb{Output}:  $\pmb d^{(a)},~ t^{(a)},~ \pmb p^{(a)}$, and Lagrange variables $\lambda^{(a)}$, $\pmb \mu^{(a)}$. By solving Problem (\ref{time}), the optimal $\pmb r^{(a)}$ and decoding order can be obtained.
   \end{algorithmic}
\end{algorithm}
\begin{algorithm}[h]
	\caption{:Algorithm for Solving Problem (\ref{sys1mindp44})}
	\label{alg:Framwork3}
	\begin{algorithmic}[1]
		\STATE \pmb{Initialize} $\pmb p^{(a)}$.
        \STATE Use the interior point method to update $\pmb d^{(a)}$, $t^{(a)},~\pmb r^{(a)},~ \pmb u^{(a)}$ by solving Problem (\ref{sys1mindp44}).
        \STATE Update $\pmb \theta^{(a)}$ based on $\pmb u^{(a)}$.
        \STATE \pmb{Output}:  $\pmb d^{(a)},~t^{(a)},~\pmb r^{(a)},~\pmb \theta^{(a)}$.
	\end{algorithmic}
\end{algorithm}
\section{TDMA Scheme for comparison}
In this section, we consider an RIS-based TDMA-MEC model to serve as a benchmark for the aforementioned RIS-based NOMA-MEC transmission model.
In the TDMA protocol, the time sharing factor of each user is $1/K$, and the rate of the $k$-th user is
{\begin{align}\label{TDMArate}
r_k = \frac{B}{K}\log_2\left( 1+ \frac{ \left\|\pmb h_{D,k}^H+\pmb h_{k}^H \pmb \Theta \pmb G\right\|^2}{\sigma^2}  {p_k} \right), \forall k\in \mathcal K.
\end{align}}
Similar to the NOMA-based model in Section II, we provide the following energy minimization problem for RIS-assisted TDMA-MEC system model
\begin{subequations}\label{sys1min_TDMA}
	\begin{align}
	\mathop{\min}_{\pmb d,  t, \pmb p, \pmb r, \pmb \theta}\quad
	&\sum_{k=1}^K (\alpha\frac{(R_{k}-d_{k})^3 C^3_{k}}{T^2}+tp_k)
	\\
	\textrm{s.t.}\quad
	& \sum_{k=1}^K d_{k} C_{k}\leq F,\\
	& tr_k\geq d_k, \quad \forall k\in\mathcal K,\\
    &{r_k \leq \frac{B}{K}\log_2\left( 1+ \frac{ \left\|\pmb h_{D,k}^H +\pmb h_{k}^H \pmb \Theta \pmb G\right\|^2}{\sigma^2}  {p_k} \right), \forall k\in\mathcal K},\\
	& t \leq T,\\
	& 0 \leq p_k \leq P_k,\quad \forall k\in\mathcal K,\\
	&  D_{k} \leq d_{k} \leq R_{k}, \quad \forall k\in\mathcal K.
	\end{align}
\end{subequations}
We use (\ref{sys1min_TDMA}d) to replace (\ref{TDMArate}) because if (\ref{sys1min_TDMA}d) holds with inequality, we can always find a smaller $p_k$ until the optimal solution is obtained. Thus (\ref{sys1min_TDMA}d) is equivalent to (\ref{TDMArate}) in Problem (\ref{sys1min_TDMA}).
It is noted that Problem (\ref{sys1min_TDMA}) is a simplified version of Problem (\ref{sys1min1}). It is because without NOMA technique, the number and complexity of TDMA rate constraints are dramatically reduced.
For solving Problem (\ref{sys1min_TDMA}), we firstly solve this problem for given $\pmb \theta$ and then use alternating 1-D search method to obtain the sub-optimal $\pmb \theta$. Then alternate the two processes until the suboptimal solution is obtained.
\subsection{Solving Problem (\ref{sys1min_TDMA}) for Given $\pmb \theta$}
It is not difficult to find that for given $\pmb \theta$, Problem (\ref{sys1min_TDMA}) can be written as follows
\begin{subequations}\label{sys1min_TDMA2}
	\begin{align}
	\mathop{\min}_{\pmb d,  t, \pmb p, \pmb r}\quad
	&\sum_{k=1}^K (\alpha\frac{(R_{k}-d_{k})^3 C^3_{k}}{T^2}+tp_k)
	\\
	\textrm{s.t.}\quad
	&(\ref{sys1min_TDMA}b)-(\ref{sys1min_TDMA}g).
	\end{align}
\end{subequations}

Problem (\ref{sys1min_TDMA2}) is a convex problem because constraint (\ref{sys1min_TDMA}c) is quasi-concave and other constraints are convex. The partial Lagrangian of Problem (\ref{sys1min_TDMA2}) is
{\begin{align}\label{lagarangian22}
&\mathcal L_{TDMA}(\pmb d,t,\pmb p,\pmb r,\upsilon,\pmb \xi,\pmb \eta)= \sum_{k=1}^K (\alpha\frac{(R_{k}-d_{k})^3 C^3_{k}}{T^2}+tp_k)\\ \nonumber
&+\upsilon( \sum_{k=1}^K d_kC_k-F)+\sum_{k=1}^K\xi_k(d_k-tr_k)\\\nonumber
&+\sum_{k=1}^K\eta_k\left(r_k-\frac{B}{K}\log_2\left( 1+ \frac{ \left\|\pmb h_{D,k}^H +\pmb h_{k}^H \pmb \Theta \pmb G\right\|^2}{\sigma^2}  {p_k} \right)\right).
\end{align}}
Correspondingly, the Lagrange dual function is
\begin{subequations}\label{sys1min3TDMA}
	\begin{align}
	g_{TDMA}(\upsilon,\pmb \xi,\pmb \eta)=\mathop{\min}_{\pmb d, t, \pmb p, \pmb r}& \mathcal L_{TDMA}(\pmb d,  t, \pmb p, \pmb r, \lambda,\pmb \mu)\\
	s.t. \quad & t \leq T,\\
	&  D_{k} \leq d_{k} \leq R_{k},~ \forall k\in\mathcal K,\\
	& 0 \leq p_k \leq P_k, ~ \forall k\in\mathcal K,
	\end{align}
\end{subequations}
where $\upsilon,~\xi_k,~ \eta_k$ are the Lagrange dual variables. The Lagrange dual problem is
\begin{subequations}\label{sys1min4TDMA}
	\begin{align}
	\mathop{\max}_{\upsilon,\pmb \xi,\pmb \eta}& \quad g_{TDMA}(\upsilon,\pmb \xi,\pmb \eta)\\
	s.t.
	&\quad  \upsilon \geq 0,\\
	&\quad  \xi_k \geq 0, \quad \forall k\in\mathcal K,\\
	&\quad  \eta_k \geq 0, \quad \forall k\in\mathcal K.
	\end{align}
\end{subequations}

To solve Problem (\ref{sys1min_TDMA2}), we obtain the optimal $\pmb d^*,~ t^*,~ \pmb p^*,~ \pmb r^*$ with given Lagrange dual variables $\upsilon,~\pmb \xi,~\pmb \eta$ by solving Problem (\ref{sys1min3TDMA}) and then update Lagrange dual variables.

\subsubsection{Evaluating $g_{TDMA}(\upsilon,\pmb \xi,\pmb \eta)$ by solving Problem (\ref{sys1min3TDMA})}
First, for given $\upsilon,~\pmb \xi,~\pmb \eta$, we obtain the dual function $g_{TDMA}(\upsilon,\pmb \xi,\pmb \eta)$ by solving Problem (\ref{sys1min3}). The Lagrangian in (\ref{lagarangian22}) can be rewritten as follows

{\begin{align}\label{lagarangian222}
&\mathcal L_{TDMA}(\pmb d,t,\pmb p,\pmb r,\upsilon,\pmb \xi,\pmb \eta)=\\ \nonumber
&\sum_{k=1}^K\left( \left(\xi_k+\upsilon C_k \right)d_{k}+\frac{\alpha(R_{k}-d_{k})^3 C^3_{k}}{T^2}\right)\\ \nonumber
&+t\left(\sum_{k=1}^K p_k-\sum_{k=1}^K\xi_kr_k\right)
+\sum_{k=1}^K\eta_kr_k\\ \nonumber
&-\sum_{k=1}^K\eta_k\frac{B}{K}\log_2\left( 1+ \frac{ \left\|\pmb h_{D,k}^H+\pmb h_{k}^H \pmb \Theta \pmb G\right\|^2}{\sigma^2}  {p_k} \right)
-\upsilon F.
\end{align}}
It is obvious that the objective function and all the constraints of Problem (\ref{sys1min3TDMA}) are decoupled for $\pmb d$ and other variables. By dropping the constant terms $-\upsilon F$, Problem (\ref{sys1min3TDMA}) is equivalent to the following Problem (\ref{p11}) and Problem (\ref{p22}), and each of them can be decomposed into $K$ subproblems:
\begin{subequations}\label{p11}
	\begin{align}
	\mathop{\min}_{\pmb d} \quad &\sum_{k=1}^K \left(\upsilon d_kC_k+\frac{\alpha \left(R_{k}-d_{k}\right)^3C^3_k}{T^2}+\xi_kd_k\right)\\
	s.t. \quad  & D_k \leq d_k \leq R_k,
	\end{align}
\end{subequations}
and
{\begin{subequations}\label{p22}
	\begin{align}
	\mathop{\min}_{\pmb r,\pmb p, t}\quad  &t\left(\sum_{k=1}^K p_k-\sum_{k=1}^K\xi_kr_k\right)+\sum_{k=1}^K\eta_kr_k\\ \nonumber
	&-\sum_{k=1}^K\eta_k\frac{B}{K}\log_2\left( 1+ \frac{ \left\|\pmb h_{D,k}^H +\pmb h_{k}^H \pmb \Theta \pmb G\right\|^2}{\sigma^2}  {p_k} \right)\\
	s.t. \quad
	& 0 \leq t \leq T,\\
	& 0 \leq p_k \leq P_k,\\
	& (\ref{sys1min_TDMA}d).
	\end{align}
\end{subequations}}
The following theorem gives the optimal solution of Problem (\ref{p11}).
\begin{theorem}
	The optimal solution of Problem (\ref{p11}) is
	\begin{align}
	d^*_k=\max \{R_k-\sqrt{\frac{\upsilon C_k+\xi_k}{3\alpha C_k}}\frac{T}{C_k}, D_k\}, \forall k\in \mathcal K.
	\end{align}
\end{theorem}
The proof of Theorem 4 is similar to that of Theorem 2.

As for Problem (\ref{p22}), if $p_k$ and $r_k$ are given, when $p_k-\xi_kr_k>0$, $t^*=0$, which means that user $k$ does not need to offload  the task to BS, and thus we can set $p^*_k=r^*_k=0$. When $p_k-\xi_kr_k\geq 0$, $t^*=T$, at this time  Problem (\ref{p22}) can be rewritten as
{\begin{subequations}\label{p222}
	\begin{align}
	\mathop{\min}_{\pmb r,\pmb p}\quad  &T\sum_{k=1}^K p_k+\sum_{k=1}^K(\eta_k-T\xi_k)r_k\\ \nonumber
	&-\sum_{k=1}^K\eta_k\frac{B}{K}\log_2\left( 1+ \frac{ \left\|\pmb h_{D,k}^H +\pmb h_{k}^H \pmb \Theta \pmb G \right\|^2}{\sigma^2}  {p_k} \right)\\
	s.t. \quad
	& 0 \leq p_k \leq P_k,\\
	& (\ref{sys1min_TDMA}d).
	\end{align}
\end{subequations}}
Define $f(\pmb r,\pmb p)$ as the objective function of Problem (\ref{p222}), i.e.,
{\begin{align}
f(\pmb r,\pmb p)=&T\sum_{k=1}^K p_k+\sum_{k=1}^K(\eta_k-T\xi_k)r_k\\ \nonumber &-\sum_{k=1}^K\eta_k\frac{B}{K}\log_2\left( 1+ \frac{ \left\|\pmb h_{D,k}^H +\pmb h_{k}^H \pmb \Theta \pmb G \right\|^2}{\sigma^2}  {p_k} \right).
\end{align}}
It is not difficult to find that $f(\pmb r,\pmb p)$ is a convex function with respect to $\pmb r$ and $\pmb p$. According to KKT conditions, we have
\begin{align}
\frac{\partial f}{\partial p_k}= T-	\frac{\eta_kB\gamma}{K(1+\gamma p_k)\ln 2}=0,
\end{align}
where
{\begin{align}
\gamma= \frac{\left\|\pmb h_{D,k}^H +\pmb h_{k}^H \pmb \Theta \pmb G \right\|^2}{\sigma^2}.
\end{align}}
Therefore, the optimal $p^*_k$ is
\begin{align}\label{p}
p^*_k=\min \left\{\frac{\eta_kB}{TK\ln 2}-\frac{1}{\gamma}, P_k\right\}.
\end{align} 
Since Problem (\ref{p222}) is a linear function with respect to $r_k$, thus when $\eta_k>T\xi_k$, the optimal $r^*_k=0$ and when $\eta_k\leq T\xi_k$, the optimal $r^*_k$ is
{\begin{align}\label{r}
r^*_k=\frac{B}{K}\log_2\left( 1+ \frac{\left\|\pmb h_{D,k}^H +\pmb h_{k}^H \pmb \Theta \pmb G \right\|^2}{\sigma^2}  {p^*_k} \right).
\end{align}}

\subsubsection{Obtaining the optimal $\upsilon^*,\pmb \xi^*,\pmb \eta^*$ by solving Problem (\ref{sys1min4TDMA})}
Since $g(\upsilon,\pmb \xi,\pmb \eta)$ is convex with respect to $\upsilon,\pmb \xi,\pmb \eta$, the Lagrange variables can be updated as follows

\begin{align}\label{l3}
\upsilon(t+1)=\left[\upsilon(t)+\delta_3(t)F-\sum_{k=1}^K d^*_kC_k  \right]^+,
\end{align}

\begin{align}\label{l4}
\xi_k(t+1)=\left[\xi_k(t)+\delta_4(t)t^*r^*_k- d^*_k  \right]^+,
\end{align}

{\begin{align}\label{l5}
&\eta_k(t+1)=\\\nonumber
&\left[\eta_k(t)+\delta_5(t)\frac{B}{K}\log_2\left( 1+ \frac{\left\|\pmb h_{D,k}^H +\pmb h_{k}^H \pmb \Theta \pmb G \right\|^2}{\sigma^2}  {p^*_k} \right)-r^*_k \right]^+,
\end{align}}
where $\delta_3(t), \delta_4(t), \delta_5(t)$ are the updating step sizes based on suitable estimates.

\subsection{Using Alternating 1-D search method to optimize $\pmb \theta$}
Since it is a benchmark scheme, we do not need to use a low complexity algorithm to guarantee the time-efficiency like Section III. In this subsection, we consider to use the alternating 1-D search method to search a sub-optimal $\pmb\theta$.

More specifically, $\pmb \Theta =\text{diag} (e^{i\theta_{1}}, \cdots, e^{i\theta_{N}})\in\mathbb C^{N\times  N}$ is a diagonal matrix with $N$ variables to be optimized, and we aim to optimize $\theta_1, \theta_2,\dots, \theta_N$ alternately. For given $\theta_2,\dots, \theta_N$, we can use the 1-D search method to obtain the optimal $\theta_1\in [0,2\pi]$ by solving Problem (\ref{sys1min_TDMA}). Similarly, we can exhaustively search $\theta_2$ with given $\theta_1,\theta_3\dots, \theta_N$ and then orderly obtain the optimal $\theta_3,\dots, \theta_N$.
The alternating iteration will continue until the difference between the results of the two iterations is less than the allowable error tolerance.
Finally, within the allowed error limit, we can alternately obtain the optimal $\theta_1,\dots, \theta_N$. This algorithm is shown in Algorithm 4.

\begin{algorithm}[h]
	\caption{:Alternating Procedure for Solving Problem (\ref{sys1min_TDMA})}
	\label{alg:Framwork2}
	\begin{algorithmic}[1]
		\STATE \pmb{Initialize} $\pmb d^{(0)}$,~ $t^{(0)}$,~ $\pmb p^{(0)}$,~ $\pmb r^{(0)}$,~ $\pmb \theta^{(0)}$, the tolerance $\tau$, iteration variable $x$.
		\REPEAT
		\STATE Update $\pmb \theta$ using alternating 1-D search method.
		\STATE $x=0,~ L(0)=0$.
		\REPEAT
		\STATE Update Lagrange variables $\upsilon,\pmb \xi,\pmb \eta$ using Eq. (\ref{l3}), Eq. (\ref{l4})  and Eq. (\ref{l5}).
		\STATE Update $\pmb p$ and $\pmb r$ according to Eq. (\ref{p}) and Eq. (\ref{r}), respectively.
		\STATE Update $\pmb d$ and $t$ by solving Problem (\ref{p11}) and (\ref{p22}).
		\STATE Calculate the objective function of Problem (\ref{sys1min_TDMA}) and the value denoted by $L(x)$, $x=x+1$.
		\UNTIL $L(x)-L(x-1)<\tau$.
		\UNTIL Alternating 1-D search method finishes.
		\STATE \pmb{Output}: the optimal $\pmb d^*,~ t^*,~ \pmb p^*,~ \pmb \theta^*,~ \pmb r^*$ , and the optimal Lagrange variables $\upsilon^*,~\pmb \xi^*,~\pmb \eta^*$.
	\end{algorithmic}
\end{algorithm}

\section{Numerical Results}
{In this section, we provide the numerical results to validate the efficiency of our proposed algorithms. We consider a two dimensional (2-D) coordinate system where the BS is located at the origin (i.e., (0 m, 0 m)) and the RIS is located at ($700$ m, $200$ m). Denote the number of reflecting elements of the RIS by $N$. For the simulation, we deploy 4 users that are randomly distributed in a $100$ m$\times100$ m square region, where the center of this area is ($1400$ m, $0$ m). The channel path loss models between BS and RIS, and from users to RIS both are $128+37.6\log_{10} d(\rm{km})$. Since the direct path is in poor condition, the direct path loss between users and the BS is $128+45\log_{10} d(\rm{km})$. The small-scale fading coefficients are assumed to be independent and identical complex Gaussian distributions with zero-mean and unit-variance \cite{chen2019FLwireless}.}
The standard deviation of path shadow fading is 8 dB, and the noise power spectral density is -174 dBm/Hz. The effective capacitance coefficient is $\alpha=10^{-28}$, the CPU cycle for computing per bit data is $C_k=10^3$ cycles/bit, and the computation offloading bandwidth is $B=1$ MHz. For convenience, we set all the computation data at each user as the same size and the maximum transmit power of each user is $P_1=P_2=P_3=1$ W. The computation capacity of user $k$ is  $F_{k}=10$ GHz $(\forall k\in\mathcal K )$  and of edge cloud is $F=20$ GHz.

In order to show the performance of our proposed RIS-based NOMA-MEC partial offloading scheme, we also compare it with other two benchmark schemes:

1) Full offloading scheme: all users are supposed to offload their computing tasks to the BS using an NOMA protocol via the RIS channel. In this case, since the maximum power constraints of users and the maximum frequency of MEC server may limit the full offloading, we assume that the MEC have the ability to process all the users' tasks and reset $F= 50$ GHz. In addition, we do not set the maximum power on all three users to make sure that they can transmit all their tasks to the MEC server.

2) Full local computing scheme: all users are supposed to compute their tasks by themselves. In this case, we just need to calculate the following local energy consumption
\begin{align}
E_{local}=\sum_{k\in\mathcal K}\alpha\frac{R_{k}^3 C^3_{k}}{T^2},
\end{align}
since there is no transmission between users and the BS.

First, we assume that the input computation data of three users are all equal to $R_k=1$ Mbits and the maximal latency of users is $T=0.6$ s. As show in Fig. 2, we illustrate the convergence behavior of NOMA-MEC scheme (scheme 1) and TDMA-MEC scheme (scheme 2) under different numbers of reflecting units $N$ at the RIS. It is obvious that the proposed algorithms converge within 7 iterations. The two transmission schemes both converge rapidly, which indicates the efficiency of proposed algorithms.
For each curve, the initial solution is high, which is because at the initial of optimization process, the phases of RIS are randomly initialized and the channels between users and BS can be in a bad condition.
After the iterations, RIS phases are optimized and the channel conditions are suitable for efficient transmission.

\begin{figure}[htb]
	\centering
	\includegraphics[width=0.5\textwidth]{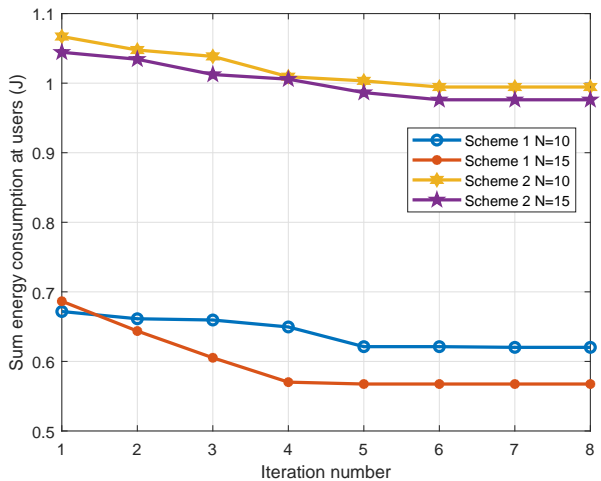}
	\vspace{-1.5em}
	\caption{Convergence behaviours of scheme 1 and 2 under different number of RIS elements $N$.}\label{fig_1}
	\vspace{-.5em}
\end{figure}

\begin{figure}[htb]
	\centering
	\includegraphics[width=0.5\textwidth]{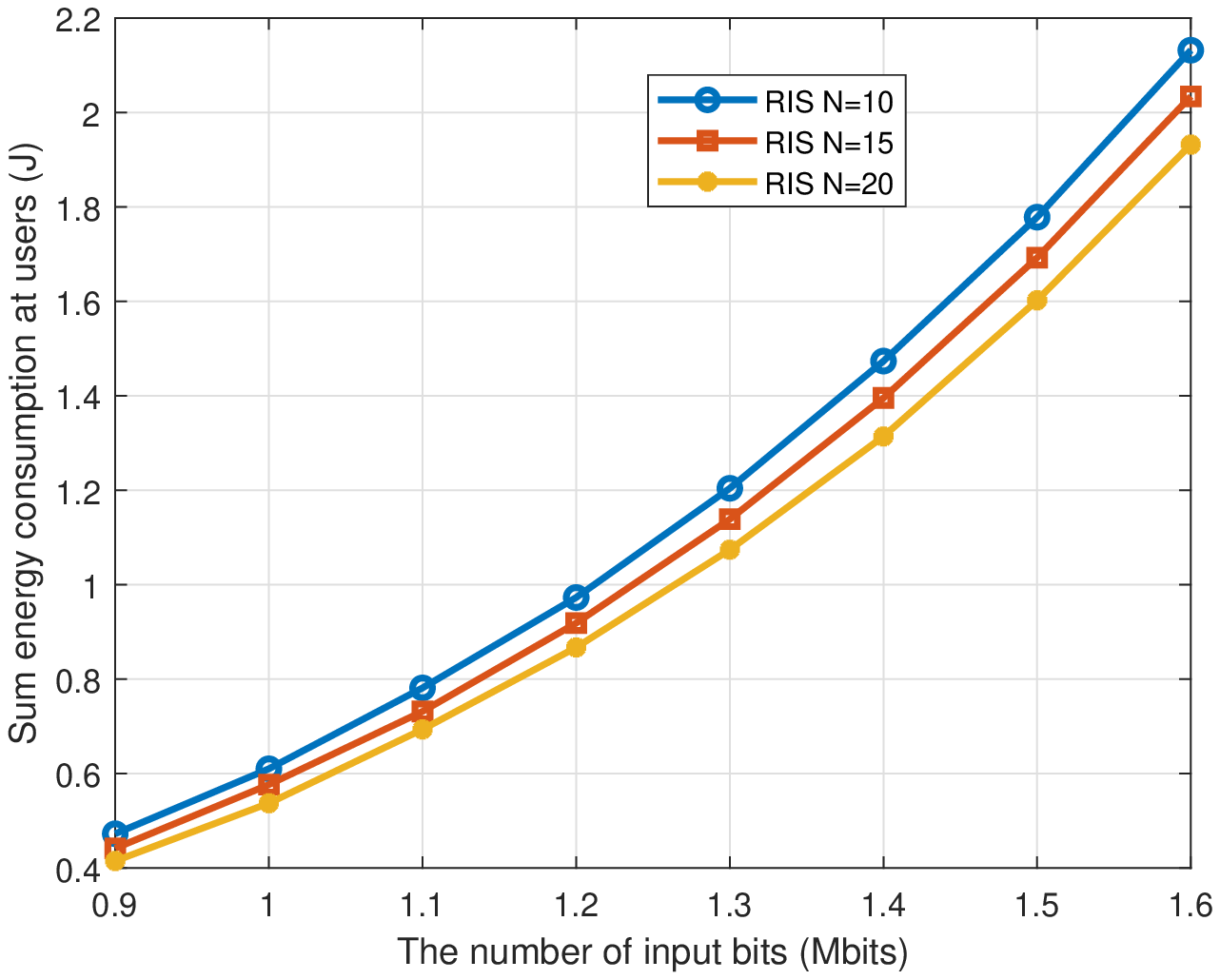}
	\vspace{-1.5em}
	\caption{Sum energy consumption versus the number of input bits.}\label{fig_2}
	\vspace{-.5em}
\end{figure}

\begin{figure}[htb]
	\centering
	\includegraphics[width=0.5\textwidth]{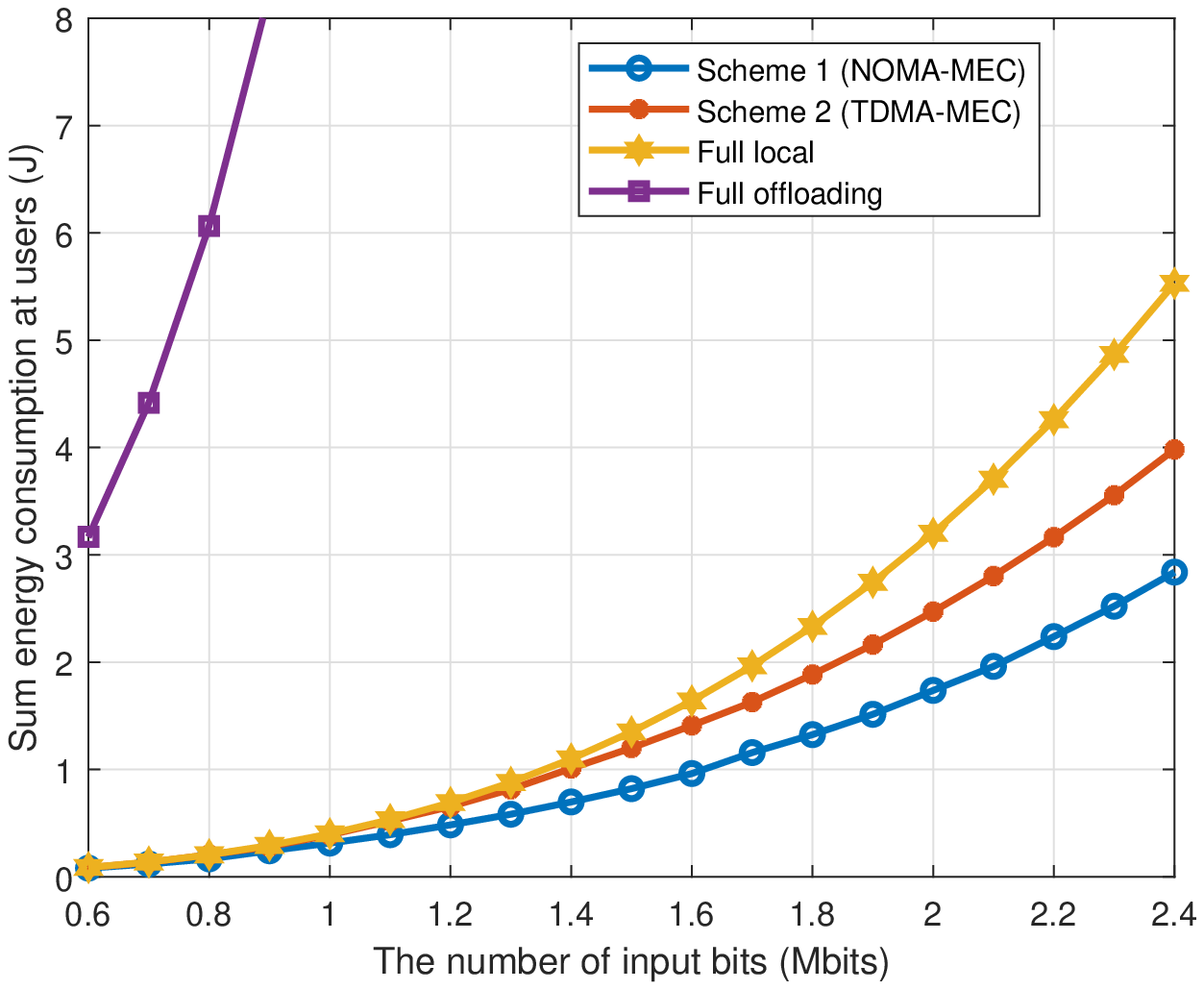}
	\vspace{-1.5em}
	\caption{ Sum energy consumption versus the number of input bits.}\label{fig_3}
	\vspace{-.5em}
\end{figure}

\begin{figure}[htb]
	\centering
	\includegraphics[width=0.5\textwidth]{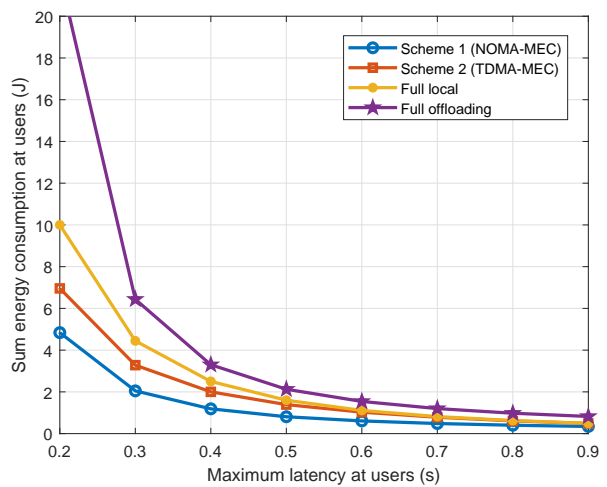}
	\vspace{-1.5em}
	\caption{ Sum energy consumption versus the maximum user latency $T$}\label{fig_4}
	\vspace{-.5em}
\end{figure}

\begin{figure}[htb]
	\centering
	\includegraphics[width=0.5\textwidth]{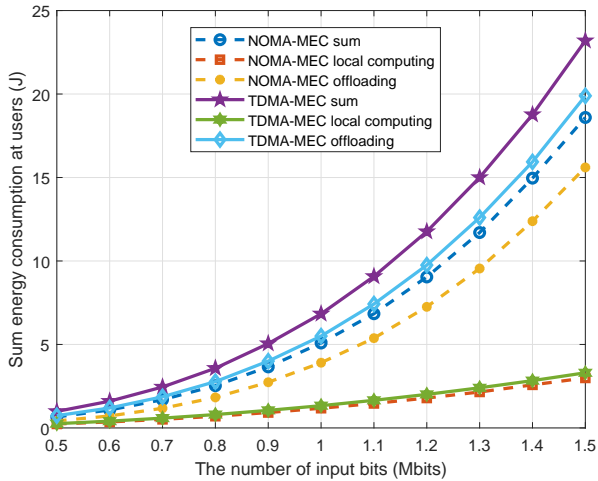}
	\vspace{-1.5em}
	\caption{ Local, offloading and sum Energy consumption versus the number of input bits.}\label{fig_5}
	\vspace{-.5em}
\end{figure}

{Fig. 3 shows the sum energy consumption at users based on the NOMA-MEC scheme versus the number of input computation tasks $R_k$, when the numbers of RIS reflecting units are $N=10,15,20$, respectively and we set the maximum user latency as $T=0.6$ s. From Fig. 3, we can see that the sum energy consumption decreases with the number of RIS units. It is because more RIS elements can improve the channel conditions between the BS and users. It is also found that the sum energy increases as the number of input bits. This is due to the fact that with the maximum latency, local users need to use more power to calculate and transmit the increasing input tasks.}

Fig. 4 compares the performance of our proposed scheme 1 with three benchmark schemes: TDMA-MEC transmission scheme (scheme 2), full offloading scheme and full local computing scheme. We set the number of RIS reflecting units as $N=15$ and the maximum latency as $T=1$ s, respectively. Numerical results illustrate that with the growing number of computation tasks, these four schemes experience different degrees of growth. Since using the full offloading scheme needs a large transmission rate to offload all the tasks to the BS, its energy consumption increases significantly.
It is also found that the energy consumption of scheme 1 achieves the best performance, especially for large input data.

In Fig. 5, we show the sum network energy consumption versus the maximum user latency $T$, where $R_k=1$ Mbits and $N=10$.
For all schemes, the energy consumption decreases with the maximum latency. The reason is that as the latency becomes longer, more computation bits can be offloaded to the MEC server, which leads to less local computation energy.
When the latency is long, the four curves grow closer, which means if the users do not care about latency, the performance of the four schemes is close. However, in a low delay system, the proposed scheme can gain a much better performance compared with other three schemes.

Fig. 6 shows the offloading energy, the local computing energy and the sum energy consumptions of NOMA-MEC and TDMA-MEC schemes versus the number of input bits, respectively. We set $T=0.2$ s and $N=10$.
It can be found that both offloading and local computing energy consumptions increase with the number of input bits and the offloading energy consumptions of the two schemes are very close.
On the other hand, the NOMA-MEC scheme consumes much lower local computing energy than the TDMA-MEC scheme, since the NOMA-MEC scheme can transmit more data bits to the BS and alleviate the local computing pressure.

{In order to show the influence of distances on the RIS-aided network, Fig. 7 illustrates the sum energy consumption at all users versus the horizontal ordinate of the RIS. Besides, to investigate the influence, we rearrange the location of four users in a square area and the coordinates of users are ($1425$ m, $25$ m) , ($1375$ m, $25$ m) , ($1425$ m, $-25$ m) , ($1375$ m, $-25$ m), respectively. The BS is located at ($0$ m, $0$ m) and the number of RIS reflecting units is $N=10$. The input computation task is $R_k=0.5$ Mbits and the maximum latency is $T=0.2$ s. We assume that the RIS moves from ($-100$ m, $50$ m) to ($1500$ m, $50$ m) with the step size of 100 m while keeping the y-coordinate fixed. It can be seen from Fig. 7 that when the RIS moves close to the BS or the users, the sum energy consumption at all users becomes low and when the RIS moves close to the approximate midpoint between the BS and users, the energy consumption reaches the highest value. The reason for having this trend is that by ignoring the small-scale fading, the large scale channel gain of the IRS aided link for user $k$ can be approximated by
\begin{align} \label{123}
	PL_k=128\times2+37.6\log_{10}d_{RB}+37.6\log_{10}d_{R,k},
\end{align}
where $d_{RB}$ and $d_{R,k}$ are the distances between RIS and BS and between RIS and user $k$, respectively.
The RIS is located close to the X-axis of coordinate system. For the convenience of explanation, we use the differences of X-axis as the approximate distances between RIS and BS or users, i.e., $d_{RB}\approx x_{RIS}-x_{BS}$ and $d_{R,k}\approx x_{k}-x_{RIS}$ (when $x_{BS}\leq x_{RIS}\leq x_{k}$), where $x_{RIS},x_{BS}$ and $x_{k}$ are the horizontal ordinates (i.e., X-axis) of the RIS, BS and user $k$, respectively. Thus Eq. (\ref{123}) can be approximated by
\begin{align} \label{1234}
	PL_k\approx &128\times2+37.6\log_{10}(x_{RIS}-x_{BS})\\\nonumber&+37.6\log_{10}(x_{k}-x_{RIS}).
\end{align}
Since $x_{BS}+x_{k}$ is the fixed distance between the BS and user $k$, $PL_k$ in Eq. (\ref{1234}) obtains the maximum value when $x_{RIS}-x_{BS}=x_{k}-x_{RIS}$, i.e., $x_{RIS}=(x_{BS}+x_{k})/2$, which means when the RIS arrives at the approximate midpoint between the BS and user $k$, the channel condition becomes the worst and the sum energy consumption reaches the peak value.}

\begin{figure}[htb]
	\centering
	\includegraphics[width=0.5\textwidth]{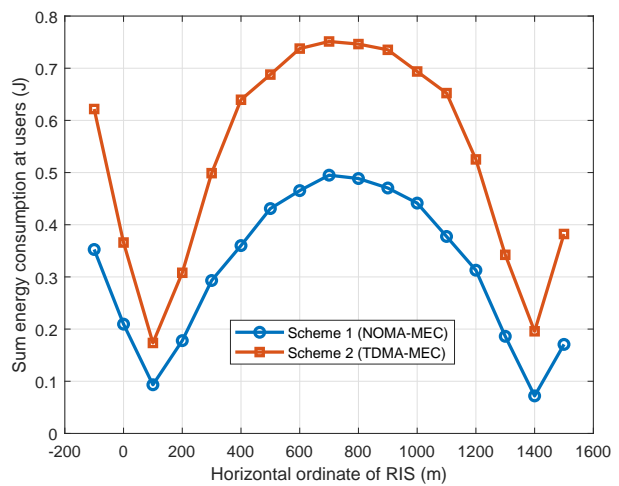}
	\vspace{-1.5em}
    \caption{ Sum energy consumption versus the horizontal ordinate of the RIS.}\label{fig_6}
	\vspace{-.5em}
\end{figure}
\begin{figure}[htb]
	\centering
	\includegraphics[width=0.5\textwidth]{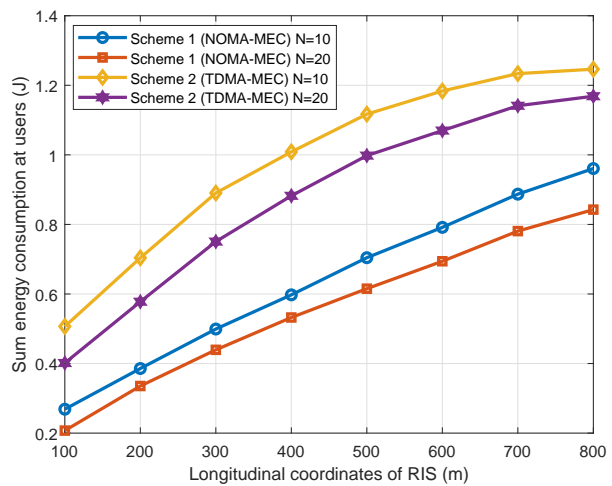}
	\vspace{-1.5em}
	\caption{ Sum energy consumption versus the longitudinal ordinate of the RIS.}\label{fig_7}
	\vspace{-.5em}
\end{figure}

{In Fig. 8, we plot the sum energy consumption versus the longitudinal ordinate of the RIS with different numbers of the RIS reflecting units. Besides, the locations of users and the BS are the same as that in Fig.7. The input computation task is $R_k=0.5$ Mbits and the maximum latency is $T=0.2$ s. We assume that the RIS moves from ($100$ m, $100$ m) to ($100$ m, $800$ m) with the step size of 100 m while keeping the x-coordinate fixed. It is obvious that the sum energy consumption at users increases with longitudinal ordinate of the RIS and the increasing speed becomes slower for large longitudinal ordinate of the RIS.}
%
%
%
%

\section{Conclusion}
In this paper, we considered an RIS-assisted NOMA-MEC network. We aimed at minimizing the energy consumption of all users by jointly optimizing the RIS phase shift, the number of transmission data, transmission rate, power control and transmission time. Since the decoding order is not determined before the transmission process and the problem is non-convex due to the RIS phase variables,
we decomposed the problem into two subproblems. For the first subproblem,
we utilized the dual method to resolve this problem with given RIS phase vector $\pmb \theta$ and obtained the closed-form solution. For the other subproblem, with given power control $\pmb p$, the penalty method is used to obtain a suboptimal solution.
The block coordinate descent process continued to alternately solve these two subproblems until the objective value converged.
Furthermore, in order to show the benefits of the proposed RIS-based NOMA-MEC scheme, we proposed three benchmark schemes, including the TDMA-MEC partial offloading scheme, full local computing scheme and full offloading scheme.
For the TDMA-MEC partial offloading scheme, we proposed an alternating 1-D search method which could easily obtain the suboptimal solution of RIS phase shifts. Numerical results showed that our proposed NOMA-MEC scheme could significantly decrease the sum energy consumptions of all users and illustrated the impact of distance on the performance of our proposed system. {In practice, the CSI might have estimation errors, and thus robust tranmission design will be studied in the future work \cite{robustcun,robustpan}.}

\bibliographystyle{IEEEtran}
\bibliography{IEEEabrv,ref}

\end{document}